\documentclass[aps, pra, reprint, nofootinbib, superscriptaddress, showpacs]{revtex4-1}
\usepackage{graphicx}
\usepackage{dcolumn}
\usepackage{bm}
\usepackage{color}
\usepackage{amsmath}
\usepackage{amssymb}

\begin{document}

\preprint{APS/123-QED}

\title{Exit momentum and instantaneous ionization rate of nonadiabatic tunneling ionization in elliptically polarized laser fields
}
\author{Siqiang Luo}
\author{Min Li}
\email{mli@hust.edu.cn} 
\author{Wenhai Xie} 
\author{Kun Liu} 
\author{Yudi Feng} 
\author{Baojie Du} 
\author{Yueming Zhou} 
\affiliation{Wuhan National Laboratory for Optoelectronics and School of Physics, Huazhong University of Science and Technology, Wuhan 430074, China} 
\author{Peixiang Lu} 
\email{lupeixiang@hust.edu.cn}
\affiliation{Wuhan National Laboratory for Optoelectronics and School of Physics, Huazhong University of Science and Technology, Wuhan 430074, China} 
\affiliation{Hubei Key Laboratory of Optical Information and Pattern Recognition, Wuhan Institute of Technology, Wuhan 430205, China}

\date{\today}

\begin{abstract}
 Based on the strong-field approximation, we obtain analytical expressions for the initial momentum at the tunnel exit and instantaneous ionization rate of tunneling ionization in elliptically polarized laser fields with arbitrary ellipticity. The tunneling electron reveals a nonzero offset of the initial momentum at the tunnel exit in the elliptically polarized laser field. We find that the transverse and longitudinal components of this momentum offset with respect to the  instantaneous field direction are directly related  to the  time derivatives of the instantaneous laser electric field along the angular and radial directions, respectively.  We further show that the nonzero initial momentum at the tunnel exit has a significant influence on  the laser phase dependence of the  instantaneous ionization rate in the nonadiabatic tunneling regime.

\end{abstract}

\maketitle

\section{introduction}

Tunneling ionization is a fundamental process in the interaction of an atom or a molecule with a strong laser pulse. An initial theoretical understanding on the tunneling ionization in an alternating electric field can be traced back to Keldysh's picture \cite{Keldysh}. This picture was extended to  elliptically polarized laser fields by Perelomov \textit{et al.} \cite{PPT,PPT2,PPT3}, which is known as Perelomov-Popov-Terent’ev (PPT) theory. Based on Keldysh's work and the PPT theory, the Ammosov-Delone-Krainov (ADK) model \cite{ADK,ADK2} is regarded as the limit when the Keldysh parameter ($\gamma=\omega \sqrt{2I_p}/\mathcal{E}$, where $\mathcal{E}$ is the laser field amplitude, $\omega$ is the field frequency, and $I_p$ is the ionization potential) is close to zero (quasistatic limit). 

The ionization rates by those pioneering works are obtained averaging over a single period of the electric field’s oscillation. To study the electron subcycle dynamics in tunneling ionization, it is  necessary  to investigate the ionization rate as a function of instantaneous laser phase. In the quasistatic limit,  the   instantaneous  ionization rate and the initial momentum at the tunnel exit can be obtained from the ADK theory. In the ADK model, the instantaneous ionization rate is a function of the instantaneous electric field and the ionization potential, but not the function of the laser frequency. Moreover, the initial momentum along the laser field direction (longitudinal momentum) at the tunnel exit is usually assumed to be zero, and the initial transverse momentum at the tunnel exit is a Gaussian distribution centered at zero. In a typical experimental condition with a Keldysh parameter of $\sim$ 1,  the ADK model becomes inaccurate. Yudin and Ivanov studied the instantaneous ionization rate including the nonadiabatic effect in a linearly polarized laser pulse  assuming a zero initial momentum of the liberated electron \cite{Yudin2001}. Based on their result, they found that the laser phase dependence of the instantaneous ionization rate in the nonadiabatic tunneling regime reveals a much broader distribution than the quasistatic limit. Bondar further derived an analytical formula for the  instantaneous ionization rate with no assumptions on the electron momentum \cite{Bondar2008}, in which the ionization rate is  a function of the electron final momentum. Recently, Li \textit{et al.} have obtained analytical expressions for the  instantaneous ionization rates as functions of the laser phase and the initial momentum at the tunnel exit \cite{Li2016}. Using different simulated methods, the instantaneous ionization rates can also be obtained \cite{Ni2016,Ivanov2018,Wangrun2019}. Those expressions of the instantaneous ionization rate are significant for the improvement of many semiclassical models in strong-field physics such as the  classical-trajectory Monte Carlo model \cite{CTMC} and quantum-trajectory Monte Carlo model \cite{Li2014,Song2016,Gong2016}. 

Tunneling ionization in elliptically polarized laser field is much more complex than that in a linearly polarized laser field because not only the magnitude but also the direction of the laser field vary rapidly \cite{Eckle2008,Eckle22008,Qin2019,Kang2018,Ma2019}. The accurate description of the instantaneous ionization rate and the initial momentum at the tunnel exit in elliptically polarized laser field is crucial for the study of attosecond angular streaking \cite{Eckle2008,Eckle22008}, nonsequential double ionization \cite{Kang2018,Ma2019}  and  elliptically polarized high-harmonic generation \cite{Fleischer2014}. 
Up to now, an analytical expression for the instantaneous ionization rate in elliptically polarized laser pulses has not been obtained.

In this paper, we derive analytical expressions for the initial momentum at the tunnel exit and the instantaneous ionization rate from tunneling ionization in elliptically polarized laser fields based on the strong-field approximation (SFA) \cite{SFA1,SFA2}. Our model including the Coulomb correction can agree well with a recent experiment \cite{Liu2019}. We prove that the initial transverse and longitudinal momentum offsets at the tunnel exit with respect to the instantaneous laser field direction are directly related to the  time derivatives of the instantaneous elliptically polarized laser field along the angular and radial directions, respectively. Because of nonzero initial momentum at the tunnel exit, the laser phase dependence of the nonadiabatic instantaneous ionization rate with the Keldysh parameter of $ \sim1 $ shows only a small difference with the quasistatic limit in the elliptically polarized laser pulse. 

\section{Subcycle nonadiabatic tunneling model}

To obtain the analytical expressions for the initial momentum at the tunnel exit and the instantaneous ionization rate, we have developed a subcycle nonadiabatic tunneling model in an elliptically polarized laser field  based on the SFA. The elliptically polarized laser field with an arbitrary ellipticity $\epsilon$ is given by the vector potential and the electric field, respectively,
\begin{eqnarray}
\begin{split}
&\textbf{A}(t)=-\frac{\mathcal{E}}{\omega}\sin(\omega t) \textbf{e}_x +\epsilon \frac{\mathcal{E}}{\omega} \cos(\omega t) \textbf{e}_y, \\
& \textbf{E}(t)=\mathcal{E}\cos(\omega t) \textbf{e}_x +\epsilon \mathcal{E} \sin(\omega t) \textbf{e}_y.
\end{split}
\end{eqnarray}

Based on the SFA involving the quantum orbits \cite{SFA1,SFA2}, the transition rate from the ground state to a continuum state $\textbf{p}=p_x\textbf{e}_x+p_y\textbf{e}_y$ can be calculated with exponential accuracy,
\begin{eqnarray}
W\propto\exp \{-2 {\rm Im}S\},
\end{eqnarray}
\begin{eqnarray}
S=-\int_{t_s}^{t_0}dt\{\frac{1}{2}[\textbf{p}+\textbf{A}(t)]^2+I_p\},
\end{eqnarray}
where $S$ is the classical action under the barrier, $t_s$ is the complex transition point, and $t_0$ is a point on the real axis of $t$ (the ionization time). 

Substituting the laser field of Eq.\,(1) into Eq.\,(3), the classical action under the barrier can be rewritten as
\begin{eqnarray}
\begin{split}
&S= i(\frac{p_{x}^{2}+p_{y}^2}{2}+I_{p}+U_{p})t_{i} \\
&-p_{x}\frac{\mathcal{E}}{\omega ^2}[\cos(\omega t_0)-\cos(\omega t_s)]  \\
& -p_y\frac{\epsilon \mathcal{E}}{\omega ^2}[\sin(\omega t_0)-\sin(\omega t_s)] \\
&+\frac{(1-\epsilon^{2}) \mathcal{E}^2}{8\omega ^3}[\sin (2\omega t_0)-\sin (2\omega t_s)],
\end{split}
\end{eqnarray}
where  $t_i$ is the  imaginary parts of the saddle-point time $t_s$  \cite{TanPRL2018,TanPRA2019} and $U_p=\frac{(1+\epsilon^2)\mathcal{E}^2}{4\omega ^2}$ is the ponderomotive energy.
The real part of Eq.\,(4) is related to a phase shift for the trajectory
under the barrier \cite{ Yan2012}, while the imaginary part is related to
the ionization rate. Here we are more interested in the ionization rate. Thus, we obtain 
\begin{eqnarray}
\begin{split}
&G\equiv {\rm Im}S=(\frac{p_{x}^{2}+p_{y}^2}{2}+I_{p}+U_{p})t_{i} \\
&- p_{x}\frac{\mathcal{E}}{\omega ^2}\sin(\omega t_0)\sinh(\omega t_i)    \\
& +p_y\frac{\epsilon \mathcal{E}}{\omega ^2}\cos(\omega t_0) \sinh(\omega t_i) \\
&- \frac{(1-\epsilon^{2}) \mathcal{E}^2}{8\omega ^3}\cos(2\omega t_0) \sinh(2\omega t_i).
\end{split}
\end{eqnarray}

Using the relation between the final canonical momentum $\textbf{p}$ and the initial momentum at the tunnel exit $\textbf{v}$, i.e., $\textbf{v}(t_0)= \textbf{p} +\textbf{A}(t_0) $, one obtains 
\begin{eqnarray}
	\begin{split}
		&p_x={\rm sgn}(\mathcal{E}_x) \bigg [ v_{||}/\sqrt{1+\epsilon^2\tan^2(\omega t_0)} \\
		&-v_{\perp}\epsilon \tan(\omega t_0)/ \sqrt{1+\epsilon^2\tan^2(\omega t_0)}   \bigg ]  +\mathcal{E}\sin (\omega t_0)/\omega \\
		&p_y={\rm sgn}(\mathcal{E}_x) \bigg [v_{||}\epsilon \tan(\omega t_0)/ \sqrt{1+\epsilon^2\tan^2(\omega t_0)} \\
		&+v_{\perp}/\sqrt{1+\epsilon^2\tan^2(\omega t_0)}\bigg ]
		-\epsilon \mathcal{E}\cos(\omega t_0)/\omega ,\\
	\end{split}
\end{eqnarray}
where $ {\rm sgn}(\mathcal{E}_x)=1  $ for $  \mathcal{E}_x >0 $, and $ {\rm sgn}(\mathcal{E}_x)=-1  $ for $  \mathcal{E}_x <0 $. $v_{||}$ and $v_{\perp}$ are the initial longitudinal and transverse momenta at the tunnel exit with respect to the instantaneous laser field, respectively. $v_{||}$ is positive when the direction is the same as the instantaneous laser field direction, and $v_{\perp}$ is positive when the direction is the same as the instantaneous rotating direction of the laser field. 

Substituting Eq.\,(6) into Eq.\,(5), we obtain that 
\begin{eqnarray}
\begin{split}
&G=[\frac{v_{||}^{2}+v_{\perp}^2}{2}+\frac{\mathcal{E}^2}{2\omega^2}(1+\epsilon^2-a^2)+I_{p}+U_{p} \\
&+\frac{\mathcal{E}v_{||}}{2a\omega}(1-\epsilon^2)\sin(2\omega t_0)-\frac{\epsilon \mathcal{E}v_{\perp}}{a\omega}]t_{i}\\
&-\frac{\mathcal{E}v_{||}}{2a\omega^2}(1-\epsilon^2)\sin (2 \omega t_0) \sinh(\omega t_i) +\frac{\epsilon\mathcal{E}v_{\perp}}{a\omega^2}\sinh(\omega t_i) \\
&-\frac{\mathcal{E}^2}{\omega^3}(1+\epsilon^2-a^2)-\frac{(1-\epsilon^{2}) \mathcal{E}^2}{8\omega ^3}\cos(2\omega t_0)\sinh(2\omega t_i),
\end{split}
\end{eqnarray}
where $a=|\mathcal{E}(t_0)|/\mathcal{E}=\sqrt{\cos^2(\omega t_0)+\epsilon^2\sin^2( \omega t_0)}$ is the normalized instantaneous laser field, and it is directly related to the laser phase when the electron exits the tunneling barrier \cite{Li2017}. $a=1$ when the ionization occurs along the major axis of the laser ellipse while $a=\lvert\epsilon\rvert$ when the ionization occurs along the minor axis.

The most probable initial transverse momentum $v_{\perp}$ at the tunnel exit at the instant of $t_0$ is determined by the condition 
\begin{eqnarray}
\partial{G}/\partial{v_{\perp}}=(v_{\perp}-\frac{\epsilon\mathcal{E}}{a \omega})t_{i}+\frac{\epsilon \mathcal{E}}{a\omega ^2}\sinh(\omega t_i)=0. \end{eqnarray}
Thus,
\begin{eqnarray}
v_{\perp}=-\frac{\epsilon \mathcal{E}}{a \omega }[\frac{\sinh (\omega t_i)}{\omega t_i}-1].
\end{eqnarray}
Further using the saddle-point equation,
\begin{eqnarray}
\frac{1}{2}[\textbf{p}+\textbf{A}(t_s)]^2+I_p=0.
\end{eqnarray}
One obtains the initial longitudinal and transverse momenta at the tunnel exit \cite{Li2017}:
\begin{eqnarray}
v_{||}=\frac{(1-\epsilon^2)\mathcal{E}\sin (2\omega t_0)}{2a\omega}[\cosh (\omega t_i)-1],
\end{eqnarray}
\begin{eqnarray}
\begin{split}
&v_{\perp}=\frac{a\mathcal{E}}{\omega} \{\sqrt{\sinh^2 (\omega t_i)-\gamma^2/a^2} \\
&-\frac{\epsilon}{a^2}[\cosh (\omega t_i)-1]\}.
\end{split}
\end{eqnarray}

Combining Eq.\,(9) and Eq.\,(12), one obtains that the imaginary time $t_i\equiv \tau/\omega$ should satisfy the transcendental equation for the most probable trajectory at the instant of $ t_0 $,
\begin{eqnarray}
\sqrt{\sinh^2 \tau-\frac{\gamma^2}{a^2}}=\frac{\epsilon}{a^2}(\cosh \tau-\frac{\sinh \tau}{\tau}).
\end{eqnarray}



Substituting Eqs.\,(11), (12), and (13) into Eq.\,(7), one obtains  the instantaneous ionization rate  with exponential accuracy,
\begin{eqnarray}
W\propto \exp[-\frac{2I_p}{\omega}f(\gamma,\epsilon,a)],
\end{eqnarray}
where
\begin{eqnarray}
\begin{split}
&f(\gamma,\epsilon,a)=[\frac{1+\epsilon^2-2a^2}{\gamma^2}\sinh^2\tau
+\frac{3(1+\epsilon^2)}{2\gamma^2}\\
&-\frac{a^2}{\gamma^2}+2]\tau 
-\frac{1}{2}(\frac{3(1+\epsilon^2)}{2\gamma^2}-\frac{a^2}{\gamma^2})\sinh(2\tau) .
\end{split}
\end{eqnarray}

The above analytical model differs from the pioneering PPT model in which only the ionization occurring along the major axis of the laser ellipse is considered \cite{PPT2}. 

Next, we establish the relation between the electron initial  momentum at the tunnel exit and the   instantaneous laser electric field.  The instantaneous angular velocity of the laser field is $\omega'=\frac{d\theta}{dt}$ with $\theta=\tan^{-1}\frac{\mathcal{E}_y}{\mathcal{E}_x}$. One can obtain the normalized instantaneous angular velocity,
\begin{eqnarray}
\alpha=\frac{\omega'}{\omega}=\frac{\epsilon}{a^2}.
\end{eqnarray}
Similarly, the instantaneous radial velocity of the laser field is $\omega''=\frac{d|\mathcal{E}(t)|}{|\mathcal{E}(t)|dt}$ with $ \lvert\mathcal{E}(t)\rvert=\sqrt{\mathcal{E}_{x}(t)^2+\mathcal{E}_{y}(t)^2} $. Thus one obtains  the normalized instantaneous radial velocity,
\begin{eqnarray}
\beta=\frac{\omega''}{\omega}=-\frac{(1-\epsilon^2) \sin(2\omega t_0)}{2a^2}.
\end{eqnarray}

Comparing Eqs.\,(16) and (17) with Eqs.\,(9) and (11), we obtain the initial longitudinal and transverse momenta at the tunnel exit for the most probable trajectory at the ionization time of $t_0$,
\begin{eqnarray}
v_{||}=-\beta \frac{a\mathcal{E}}{\omega}(\cosh \tau -1),
\end{eqnarray}
\begin{eqnarray}
v_{\perp}=-\alpha \frac{a\mathcal{E}}{\omega}(\frac{\sinh \tau}{\tau}-1).
\end{eqnarray}

From Eqs.\,(18) and (19), one knows that the longitudinal component of the initial momenum for the most probable trajectory is directly related  to the normalized instantaneous radial velocity, while its transverse component is directly related  to the normalized instantaneous angular velocity of the laser electric field in elliptically polarized laser pulses.

\section{Results and discussions}

Equations\,(13), (15), (18), and (19) are the main results of this paper. In the following, we consider some special cases in order to establish connections of this model with previous results. 

(1) In the case of $\epsilon=0$ (linear polarization), Eq.\,(13) becomes $\sinh \tau=\gamma/a$. Thus we obtain that the initial longitudinal and transverse momenta for the most probable trajectory are $v_{||}=\frac{\mathcal{E}\sin( \omega t_0)}{\omega}(\sqrt{1+\frac{\gamma^2}{a^2}}-1)$ and $v_{\perp}=0$, respectively, and Eq.\,(15) is changed to
\begin{eqnarray}
\begin{split}
&f(\gamma,\epsilon=0,a)=[\frac{1+2\sin^2 (\omega t_0)}{2\gamma^2}+\frac{1}{a^2}]\sinh^{-1}\frac{\gamma}{a}  \\ 
&-\frac{1+2\sin^2 (\omega t_0)}{2\gamma^2}\frac{\gamma}{a}\sqrt{1+\frac{\gamma}{a}^2}.
\end{split}
\end{eqnarray}
This is one of the main results in Ref.\,\cite{Li2016}, and the initial longitudinal momentum at the tunnel exit has recently been confirmed in an experiment \cite{Li2019}.

Further considering the case of $a=1$ in a linearly polarized laser field, e.g., $t_0=0$, the most probable initial momenta are $v_{||}=0$ and $v_{\perp}=0$. In this limit, we obtain the original Keldysh formula \cite{Keldysh},
\begin{eqnarray}
\begin{split}
&f(\gamma,\epsilon=0,a=1)=(\frac{1}{2\gamma^2}+1)\sinh^{-1}\gamma\\
&-\frac{\sqrt{1+\gamma^2}}{2\gamma}.
\end{split}
\end{eqnarray}

(2) In the case of $\epsilon=1$ (circular polarization), $a$ is equal to 1 for arbitrary ionization time, and thus Eq.\,(13) becomes $\sqrt{\sinh^2 \tau-\gamma^2}=\cosh \tau-\frac{\sinh \tau}{\tau}$. The most probable initial momenta are $v_{||}=0$ and $v_{\perp}=\frac{\mathcal{E}}{\omega}(\frac{\sinh \tau}{\tau}-1)$. Equation\,(15) is rewritten as
\begin{eqnarray}
f(\gamma,\epsilon=1,a=1)=\frac{2(1+\gamma^2)}{\gamma^2}\tau-\frac{1}{\gamma^2}\sinh(2\tau). 
\end{eqnarray}

(3) In the case of $a=1$ (e.g., $t_0=0$) for an arbitrary ellipticity value, $v_{||}=0$ and $v_{\perp}=\frac{\epsilon \mathcal{E}}{\omega}(\frac{\sinh \tau}{\tau}-1)$. Equation\,(15) can be rewritten as
\begin{eqnarray}
\begin{split}
	&f(\gamma,\epsilon,a=1)=(1+\frac{1+\epsilon^2}{2 \gamma^2})\tau  \\ & -\frac{1+\epsilon^2-2\epsilon^2(1-\frac{\tanh \tau}{\tau})}{2\gamma^2}\sinh\tau\cosh \tau,
\end{split}
\end{eqnarray}
where $\tau$ is determined by the transcendental equation of $\sqrt{\sinh^2 \tau-\gamma^2}=\epsilon(\cosh \tau-\frac{\sinh \tau}{\tau})$. This result is the same as the PPT model \cite{PPT2,Mur2001}.

(4) In the case of $a=\epsilon$ (e.g., $\omega t_0=\pi/2$) for an arbitrary ellipticity value, $v_{||}=0$ and $v_{\perp}=\frac{\mathcal{E}}{\omega}(\frac{\sinh \tau}{\tau}-1)$. Equation\,(15) can be rewritten as
\begin{eqnarray}
\begin{split}
&f(\gamma,\epsilon,a=\epsilon)=(1+\frac{1+\epsilon^2}{2 \gamma^2})\tau  \\ & -\frac{1+\epsilon^2-2(1-\frac{\tanh \tau}{\tau})}{2\gamma^2}\sinh\tau\cosh \tau,
\end{split}
\end{eqnarray}
where $\tau$ is determined by the equation of $\sqrt{\sinh^2 \tau-\gamma^2 /\epsilon^2}=\frac{1}{\epsilon}(\cosh \tau-\frac{\sinh \tau}{\tau})$. This is consistent with our recent work \cite{Liu2019}.

\begin{figure}
	
		\includegraphics[width=9cm]{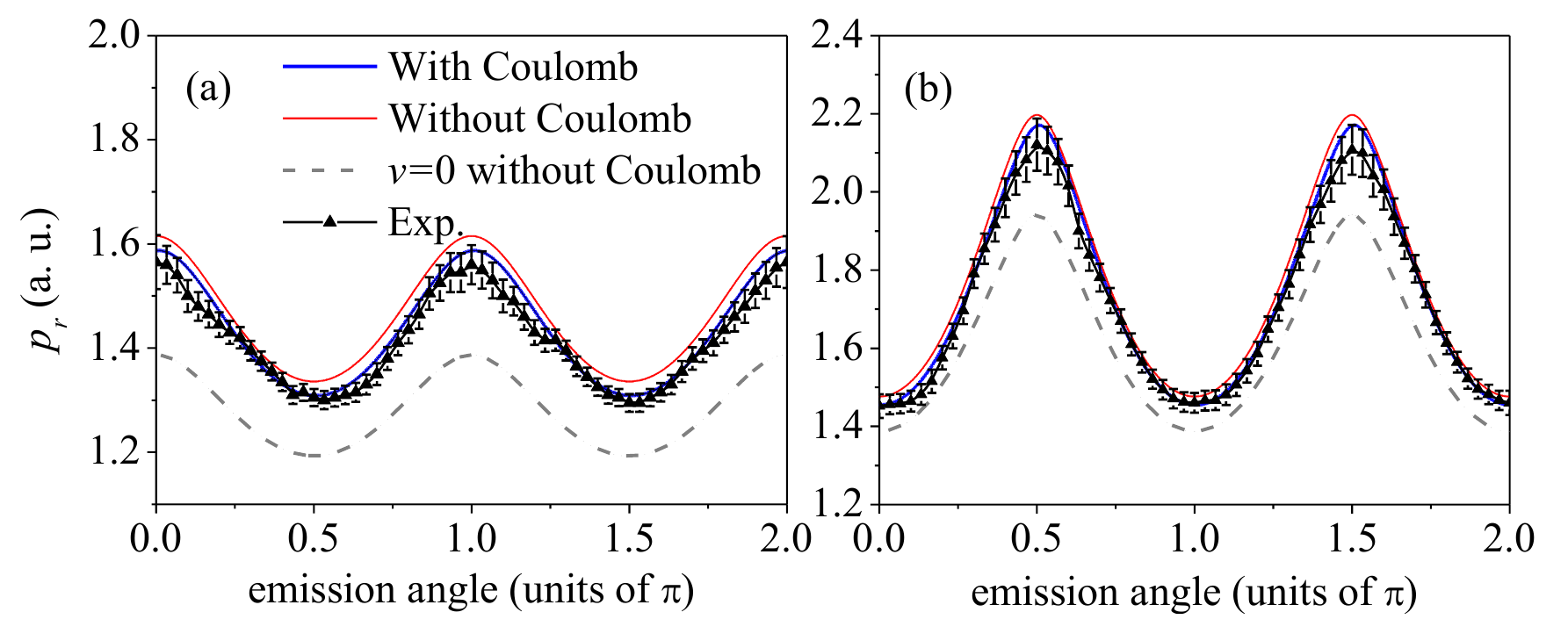}
		\caption{\label{fig1} 
			The  momentum drift with respect to the electron emission angle in elliptically polarized laser fields with the electric field ratio $E_x/E_y$ of 0.86 (a) and 1.4 (b). The  nonadiabatic results with and without including the Coulomb correction are shown by the blue and red curves, respectively. The adiabatic results with zero initial momentum at the tunnel exit are shown by the gray dashed lines. The experimental results are taken from Ref.\,\cite{Liu2019}.
		}
	
\end{figure}

To further validate the present nonadiabatic tunneling theory, we show in Fig.\,1 the momentum drift with respect to the emission angle for two ellipticities. The experimental data are taken from our recent work \cite{Liu2019}. To consider the Coulomb correction, we calculate the electron final momentum by numerically solving the classical Newtonian equation with consideration of the Coulomb potential. The initial position of the Newtonian equation is set to be the tunnel exit point, which is obtained by $ \textbf{r}(t_0 ) =
\int^{t_0}_{t_s} dt [\textbf{p} + \textbf{A}(t )] $. The final emission angle and the momentum drift are calculated by $\varphi=\tan^{-1} p_x/p_y$ and $p_r=\sqrt{p_x^2+p_y^2}$, where $ p_x $ and $ p_y $ are the final electron momenta along the major and minor axes of the laser ellipse, respectively. One can see that the nonadiabatic model including the Coulomb correction agrees well with the measurement, while the prediction of the adiabatic model is much smaller than the experiment at both ellipticity values.

\begin{figure}
	
	\includegraphics[width=9cm]{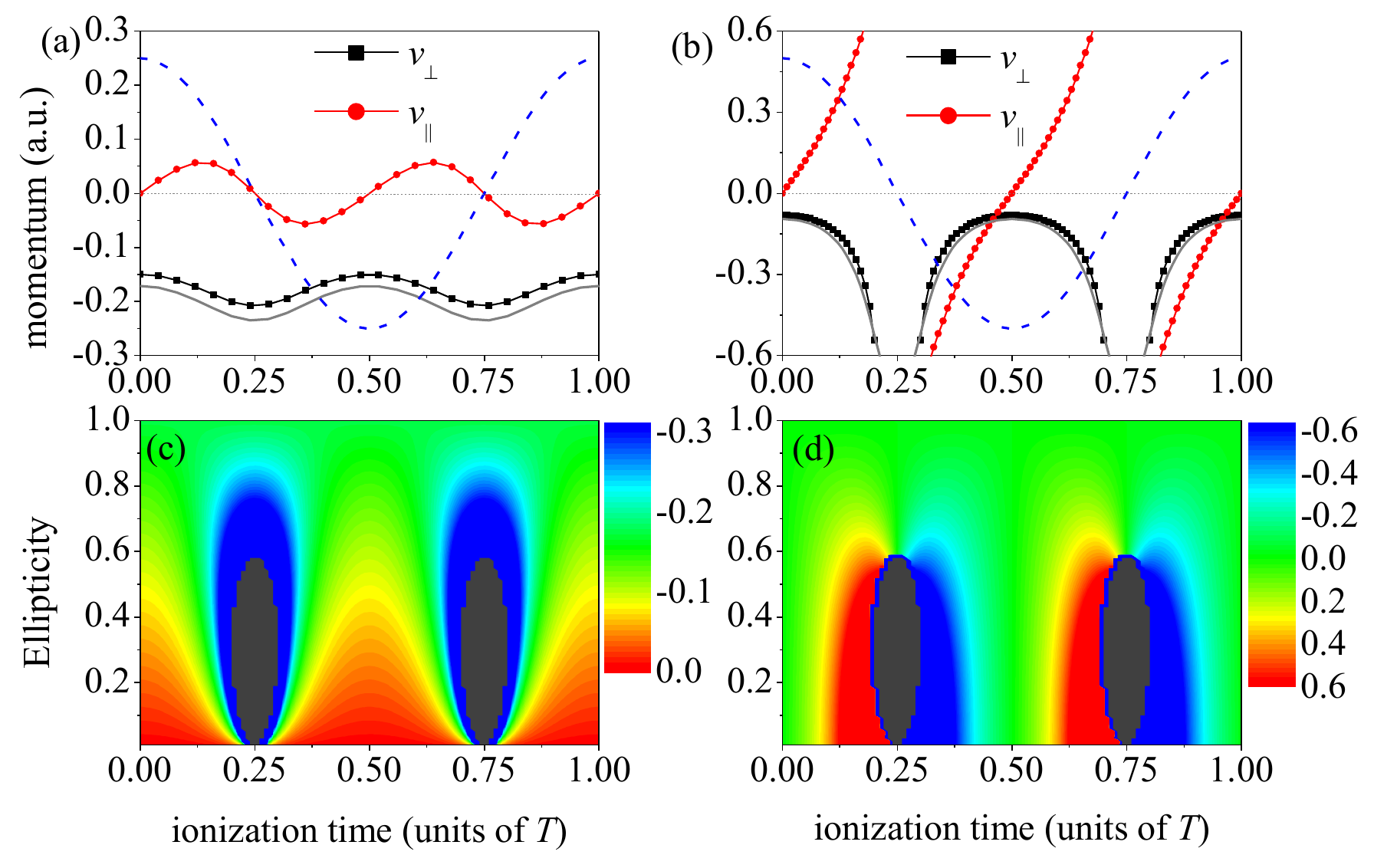}
	\caption{\label{fig2}
		(a, b) The initial transverse momentum (black curves with squares) and initial longitudinal momentum (red curves with dots) offsets at the tunnel exit for the most probable trajectory as a function of the ionization time at  ellipticities of 0.9 and 0.5, respectively. The solid gray lines show the initial transverse momentum at the tunnel exit predicted by the adiabatic theory \cite{Ohmi2015} at the corresponding ellipticities. The blue curves show the laser electric field along the $x$ direction with arbitrary units. (c, d) The most probable initial transverse and longitudinal momenta with respect to the ellipticity and the ionization time, respectively. The gray regions correspond to where we have not found the solutions for the transcendental equation of Eq.\,(13). The Keldysh parameter  is  0.91 for all ellipticities.
	}	
\end{figure}

The difference between the adiabatic and nonadiabatic results mainly comes from the effect of the initial momentum at the tunnel exit. The nonadiabic model predicts nonzero initial momentum at the tunnel exit according to Eqs.\,(18) and (19), whereas the adiabatic model assumes zero initial momentum for the most probable trajectory. Thus the initial momentum at the tunnel exit plays a crucial role on the final electron momentum distribution. In Figs.\,2(a) and 2(b), we show the initial transverse and longitudinal momenta at the tunnel exit for the most probable electron trajectory at  ellipticities of 0.9 and 0.5, respectively ($\mathcal{E}=0.79$ a.u.). At the ellipticity of 0.9, both $v_{||}$ and $v_{\perp}$ oscillate with the ionization time with an amplitude of $ \sim0.05 $ a.u. The absolute value of the initial transverse momentum is much larger than that of the initial longitudinal momentum. At the ellipticity of 0.5, the absolute value of the initial transverse momentum is very small near the field maximum and it becomes large when the ionization time approaches $0.25T$ or $0.75T$. The initial longitudinal momentum at $\epsilon=0.5$ is nearly linear with the ionization time.  

For comparison, we also show the initial transverse momentum predicted by the adiabatic theory \cite{Ohmi2015} with the solid gray lines. The adiabatic theory was developed in Ref.\,\cite{Tolstikhin2012}  for finite-range potentials and arbitrary polarization of the laser field, which amounts to the asymptotics of the solution to the time-dependent Schr\"{o}dinger equation when the adiabatic parameter approaches zero. For a long laser pulse, the adiabatic parameter is defined as $ \omega/\Delta E $ in the adiabatic theory, where $ \omega $ is the laser frequency and $ \Delta E $ is the energy spacing between the initial state and the nearest eigenstate of the atom. With further considering the first-order nonadiabatic correction in the adiabatic theory, i.e.,  Eq.\,(45) of Ref.\,\cite{Ohmi2015}, the initial transverse momentum at the tunnel exit can be expressed as $  v_{\perp}=\frac{\sqrt{2I_p}}{6}\gamma_i $, where $ \gamma_i  $ is the instantaneous effective Keldysh parameter \cite{Liu2019}. Using the laser field of Eq.\,(1), the initial transverse momentum can be rewritten as $ v_{\perp}=\frac{\epsilon\omega I_p}{3a^{3}\mathcal{E}} $ in the elliptically polarized laser field. One can see that the initial transverse momentum momenta predicted by the adiabatic theory \cite{Ohmi2015} is  close to the results predicted by our model.   
 
Figures\,2(c) and 2(d) show the initial transverse and longitudinal momenta at the tunnel exit for the most probable trajectory with respect to the ellipticity and the ionization time. The gray regions in Figs.\,2(c) and 2(d) correspond to where we have not found the solutions for the transcendental equation of Eq.\,(13). For a near linearly polarized laser field ($\epsilon<<1$), the absolute value of the initial transverse momentum is very small. With increasing the ellipticity, the absolute value of the initial transverse momentum becomes larger at the maximum of the laser field. The absolute value of the initial transverse momentum becomes much larger when the laser ellipticity is within [0.5, 0.8] at the laser minima of $t_0=0.25T$ and $t_0=0.75T$.
As shown in Fig.\,2(d), the most probable initial longitudinal momentum at the tunnel exit is zero at the laser extremum because of the vanishing instantaneous radial velocity of the laser field. When the ionization occurs away from the laser extremum, the absolute value of the initial longitudinal momentum becomes large. The most probable longitudinal momentum approaches zero for all ionization times when the laser ellipticity is very close to 1.

\begin{figure}
	\includegraphics[width=9 cm]{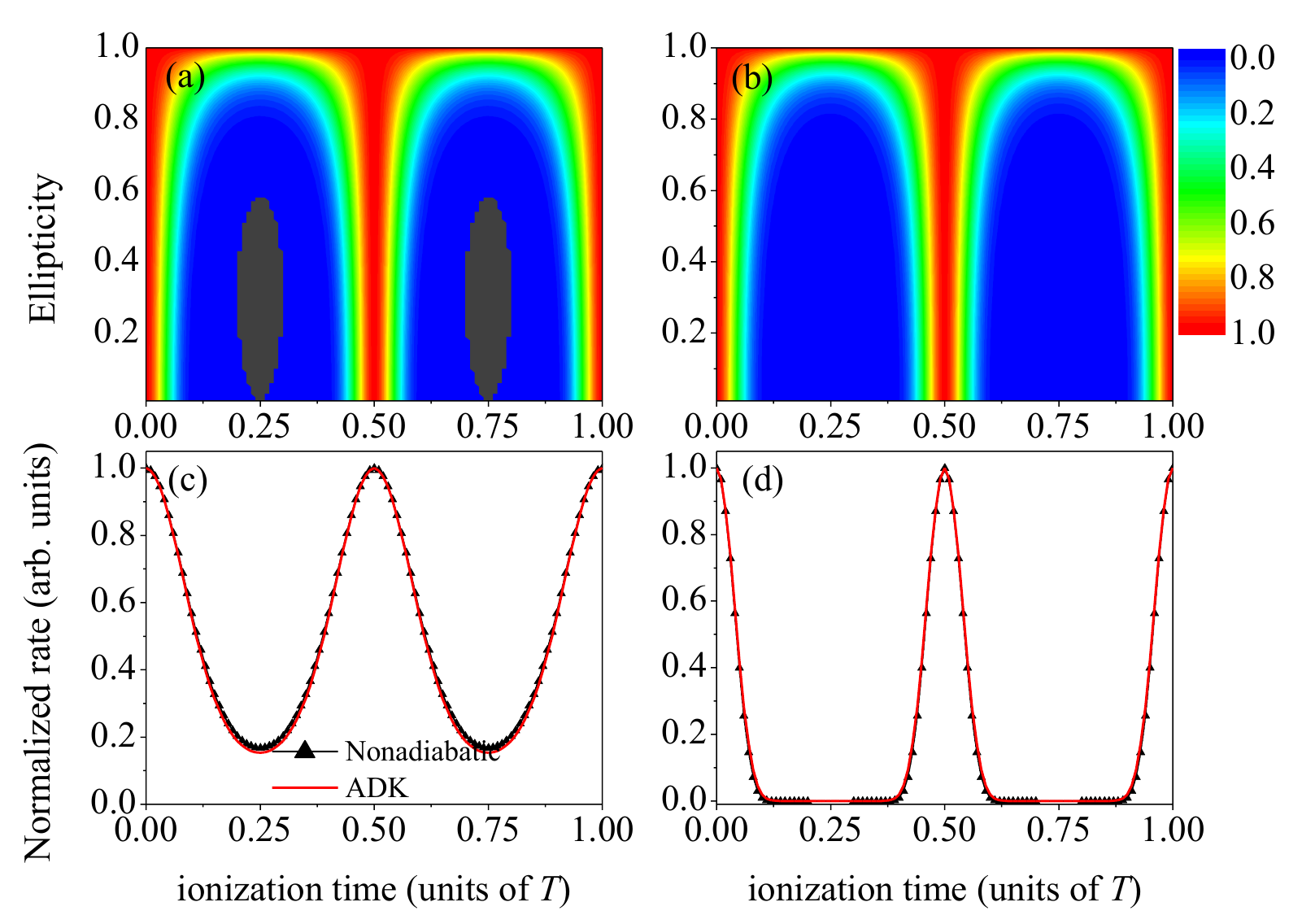}
	\caption{\label{fig3} 
		(a, b) The instantaneous ionization rate with respect to the ellipticity and the ionization time using the nonadiabatic theory and the ADK theory, respectively. (c, d) Lineouts taken at ellipticities of 0.9 and 0.1, respectively. The data are normalized to the ionization rate at the field maximum for each ellipticity. The Keldysh parameter  is  0.91 for all ellipticities.
	}	 
\end{figure}

We next concentrate on the instantaneous ionization rate by the subcycle nonadiabatic tunneling theory. Figure\,3(a) shows the instantaneous ionization rate with respect to the ellipticity and the ionization time using Eqs.\,(14) and (15). The data are normalized to the ionization rate at the field maximum for each ellipticity. For comparison, the result based on the adiabatic ADK theory \cite{ADK,ADK2} is shown in Fig.\,3(b). Here we have neglected the preexponential factor for the ADK theory. One can see that the laser phase dependence of the instantaneous ionization rate from the subcycle nonadiabatic tunneling theory is very similar to the quasistatic limit, i.e., the ADK result. The lineouts taken at the ellipticities of 0.9 and 0.1 are shown in Figs.\,3(c) and 3(d), respectively. The distribution of the normalized ionization rate is nearly the same for the nonadiabatic and adiabatic results at both ellipticities. This  is different with the result in Ref.\,\cite{Yudin2001}, where the nonadiabatic instantaneous ionization rate as a function of the laser phase  is much broader than that of the quasistatic ADK theory.

\begin{figure}
	\centering
	\includegraphics[width=9.0cm]{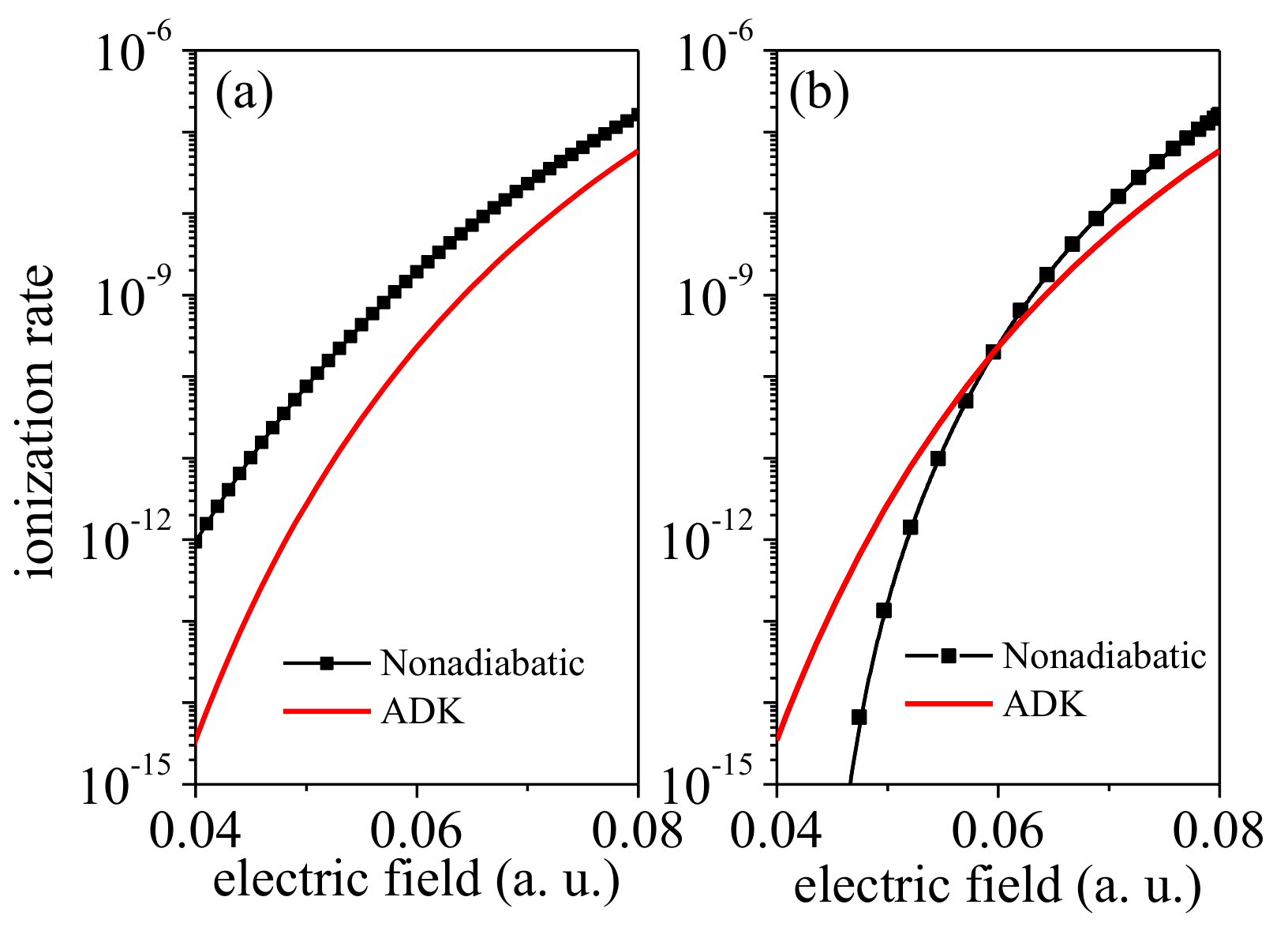}
	\caption{\label{fig4}
		(a) The ionization rate with respect to the peak laser field ($\mathcal{E}$)  with $a=1$, i.e., only the ionization  along the major axis of the laser ellipse is considered. (b) The instantaneous ionization rate with respect to the instantaneous electric field ($a\mathcal{E}$) within one quarter laser cycle for $\mathcal{E}=0.08$ a.u.. The laser ellipticity is 0.5 for both panels. Note the ionization rate is shown in logarithmic scale.
	}
\end{figure} 

The main difference of the instantaneous ionization rate between this work and Ref.\,\cite{Yudin2001}  is that we have included the effect of the initial momentum at the tunnel exit in this work. To show the effect of the initial momentum at the tunnel exit on the instantaneous ionization rate, we compare two cases in Fig.\,4 using our model. Figure\,4(a) shows the ionization rate (in logarithmic scale) with respect to the peak electric field $ \mathcal{E} $ (equivalent to the intensity) with $ a=1 $, while Fig.\,4(b) shows the ionization rate with respect to the instantaneous electric field within a quarter of cycle. In both cases, the range of the electric field ($ a\mathcal{E} $) is the same.  
Using the adiabatic tunneling theory, i.e., ADK theory, the ionization rates are the same for Figs.\,4(a) and 4(b), as shown by the red solid curves. However, the nonadiabatic ionization rate differs significantly for those two cases. As shown in Fig.\,4(a), the nonadiabatic ionization rate decreases more slowly  than the ADK result with the decrease of the peak electric field, which  is consistent with the PPT theory \cite{PPT,PPT2,PPT3}. In contrast, the instantaneous ionization rate in Fig.\,4(b) decreases much faster than the ADK theory with the decreasing of the instantaneous electric field strength. This comes from the effect of the initial momentum at the tunnel exit. With decreasing  the laser electric field strength within a quarter of the laser cycle, the electron obtains a large initial momentum at the tunnel exit. This makes it more difficult  for the electron to penetrate the tunneling barrier. As a result, the instantaneous ionization rate decreases more rapidly than expected.
Therefore, the instantaneous ionization rate as a function of the laser phase is different with the ionization rate as a function of the laser peak field strength. Considering that  the electron is mainly released near the field maximum when $\epsilon<<1$, the phase dependence of the nonadiabatic instantaneous ionization rate can be approximated by the ADK tunneling formula in the typical condition of elliptically polarized laser pulses, as shown in Fig.\,3.

\section{conclusion}

In summary,  we have derived analytical expressions for the most probable initial momentum at the tunnel exit and the instantaneous ionization rate for tunneling ionization in an elliptically polarized laser pulse with arbitrary ellipticity. We find that the initial transverse momentum for the most probable trajectory is directly related to the instantaneous angular velocity of the laser field while its initial longitudinal momentum is directly related to the instantaneous radial velocity of the laser field in the elliptically polarized laser pulse. Due to the nonzero initial momentum at the tunnel exit, the laser phase dependence of the instantaneous ionization rate shows a small difference with the quasistatic limit under typical experimental conditions in the elliptically polarized laser field. Our study concerns only the exponential dependence of the nonadiabatic tunneling process. Further including the preexponential factor taken from the PPT theory \cite{PPT,PPT2,PPT3} will make the present study more complete. Recently, it has been shown that the magnetic quantum number plays an important role in the tunneling ionization in elliptically or circularly polarized laser fields \cite{ Barth2011,Herath2012,Hartung2016,Eckart2018,LiuKunLong2018,Zhang2019}, which mainly affects the preexponential factor of the ionization rate. Thus our study will have significant applications for investigating the effect of the magnetic quantum number on tunneling ionization in elliptically or circularly polarized laser pulses. 

\section{acknowledgment}

This work was supported by the National Natural Science Foundation of China (Grants No. 11627809, 11674116, and 11722432) and Program for HUST Academic Frontier Youth Team.

\end{document}